\documentclass{aa}

\usepackage[breaklinks=true,colorlinks,citecolor=blue]{hyperref}
\usepackage{graphicx}
\usepackage{enumerate}
\usepackage{epsfig}
\usepackage{color}
\usepackage{txfonts}
\usepackage{natbib}
\usepackage{multirow}
\usepackage{ulem}
\usepackage{amssymb}
\usepackage{xcolor}

\definecolor{ao}{rgb}{0.0, 0.5, 0.0}

\newcommand{\MyTabA}{\ref{MyTabA}}
\newcommand{\MyFigA}{\ref{MyFigA}}

\newcommand{\Msun}{M_{\odot}}
\newcommand{\rate}{\rm Gpc^{-3}~yr^{-1}}
\begin{document}

\title{Constraining hierarchical mergers of binary black holes detectable \\with LIGO-Virgo}
\author{Guo-Peng Li}%{\href{https://orcid.org/0000-0003-3306-5217}
\institute{Guangxi Key Laboratory for Relativistic Astrophysics, School of Physical Science and Technology, Guangxi University, \\Nanning 530004, PR China\\
              \email{lgp@st.gxu.edu.cn}}
\date{Received 13 June 2022 / Accepted 18 August 2022}

\abstract{
Most of the binary black hole (BBH) mergers detected by LIGO and Virgo could be explained by first-generation mergers formed from the collapse of stars, while others might come from second (or higher) generation mergers, namely hierarchical mergers, with at least one of the black holes (BHs) being the remnant of a previous merger. A primary condition for the occurrence of hierarchical mergers is that the remnant BHs can be retained by the host star cluster. We present a simple formula to estimate the hierarchical merger rate in star clusters. We find this latter to be proportional to the retention probability. Further, we show that $\sim 2\%$ of BBH mergers in nuclear star clusters (NSCs) may instead be of hierarchical mergers, while the percentage in globular clusters (GCs) is only a few tenths of a percent. However, the rates of hierarchical merger in GCs and NSCs are about the same, namely of $\sim {\mathcal O(10^{-2})}~{\rm Gpc^{-3}~yr^{-1}}$, because the total BBH merger rate in GCs is larger than that in NSCs. This suggests that if a gravitational-wave event detected by LIGO-Virgo is identified as a hierarchical merger, then it is equally likely that this merger originated from a GC or an NSC.
}
\maketitle
%%%%%%%%%%%%%%%%%%%%%%%%%%%%%%%%%%%%%%%%%%%%%%%%%%%%%%%%%%%%%%%%%%%%%%%%%%%%%%%
%---------------------------------------------------------------------------------
%
\section{Introduction}\label{Sec:Intro}
Since the first gravitational-wave (GW) event, GW150914~\citep{GW150914}, detected by Advanced LIGO~\citep{LIGO-2015} and Advanced Virgo~\citep{Virgo-2015}, more than 80 stellar-mass binary black hole (BBH) mergers have been reported by the LIGO-Virgo-KAGRA (LVK) Collaboration~\citep{GWTC-1,GWTC-2,GWTC-2.1,GWTC-3}.
The two main formation channels of BBH mergers have been extensively discussed, and include isolated binary evolution and dynamical formation~\citep{Mapelli-2021hgwa.bookE...4M,Ilya-2022PhR...955....1M}.
Dynamical formation channels occur only in dense environments such as nuclear star clusters (NSCs) or globular clusters (GCs), although a body of work has shown that dynamical formation channels produce mergers in the field of the host galaxies~\citep[i.e.,][]{2019ApJ...887L..36M,2020MNRAS.498.4924M,2022MNRAS.514.4246R,2022arXiv220515040M}.
In the above classical classification, the dynamical formation channel from wide systems in the field is ignored.

Some special GW events~\citep[e.g.,][]{GWTC-1,GW190412,GW190521} with significant asymmetric masses, massive components, or high spins are promising candidates for hierarchical BBH mergers---with at least one of the black holes (BHs) being the remnant of a previous merger~\citep{Gerosa-Davide-2021-review}. In particular, GW190521~\citep{GW190521} is regarded as a promising candidate to date~\citep{Anagnostou-2020arXiv201006161A,Dall'Amico-2021MNRAS.508.3045D,Kimball-2021-ApJ.915L35,
DG-PhysRevD.104.084002,2022ApJ...927..231F}.
Hierarchical mergers can occur via dynamical formation channels ~\citep{Miller-2004IJMPD.13.1,Gerosa-Davide-PhysRevD.100.041301,Yang-Y-PhysRevLett.123.181101,
Dall'Amico2021arXiv211202020D,Li-PRD-2022-105.063006} or triple-body (and higher) configurations~\citep{Hamers_A-2020-Safarzadeh_M-ApJ.898.99,Lai-2021MNRAS.502.2049L}.
A primary condition for the occurrence of hierarchical mergers is that the remnants of first-generation BBH mergers are retained by the host star cluster (the binary also needs to survive few-body interactions).
Therefore, the escape speed of the host star cluster needs to be larger than the typical recoil velocity (or ``kick'') of remnant BHs.

Energy, angular momentum, and linear momentum are carried by GW radiation, in which the loss of linear momentum causes a kick imparted to a remnant BH.
Measuring the kick velocity is significantly important in the inference of the hierarchical merger rate.
The kick can be extracted from GW signals but with great uncertainty~\citep{Calderon-Bustillo-PhysRevLett.121.191102,Varma-Vijay-PhysRevLett.124.101104}.
The kick velocities inferred from the Gravitational Wave Transient Catalog (GWTC) events lie in a wide range:  $\sim 50{-}2\,000~\rm{km~s^{-1}}$~\citep{Mahapatra-Parthapratim-2021ApJ...918L..31M,2022PhRvL.128s1102V}.
In comparison, the escape speed of a typical GC is $\sim 2{-}100~\rm{km~s^{-1}}$ and that of an NSC is $\sim 10{-}600~\rm{km~s^{-1}}$~\citep{Antonini-2016ApJ...831..187A}.

Recently,~\cite{Fragione-2021-Loeb-MNRAS.502.3879} computed the distribution of kick velocities of the GWTC-2 events~\citep{GWTC-2} as a function of the spins of two BHs in a binary and found that remnant BHs with dimensionless spins as small as 0.1 would be retained by NSCs with escape speeds of higher than $100~\rm{km~s^{-1}}$.
\cite{Doctor-Zoheyr-2021ApJ...914L..18D} obtained the kick velocity distributions of the GWTC-2 events using a numerical relativity (NR) surrogate waveform model~\citep{Varma-Vijay-PhysRevResearch.1.033015,Varma-Vijay-PhysRevLett.122.011101} and found that GCs and NSCs may retain up to $\sim 3\%$ and $\sim 46\%$ of their remnant BHs, respectively.
\cite{Mahapatra-Parthapratim-2021ApJ...918L..31M} estimated the kick velocities of the GWTC-2 events using NR fitting formulas~\citep{Gonzalez-Jose-2007ApJ...659L...5C} and found that $\sim 50\%$ of remnant BHs can be retained by NSCs with escape speeds of $200~\rm{km~s^{-1}}$.
These results will independently constrain the rates of hierarchical mergers.

In this paper, we present a method to estimate the rates of hierarchical mergers in star clusters and give a simple formula in Sect.~\ref{Sec:Method}.
We demonstrate that the rate of hierarchical merger is approximately proportional to the retention probability.
In Sect.~\ref{Rate} we apply the formula to estimate hierarchical merger rates in GCs and NSCs.
Finally, in Sect.~\ref{Discussion} we discuss the comparison of our results  with the simulations, and we summarise our main results in Sect.~\ref{Conclusion}.
%
%%%%%%%%%%%%%%%%%%%%%%%%%%%%%%%%%%%%%%%%%%%%%%%%%%%%%%%%%%%%%%%%%%%%%%%%%%%%%%%
%---------------------------------------------------------------------------------

%
\section{Method}\label{Sec:Method}

\subsection{Formula derivation}
We assume that there are no hierarchical mergers currently taking place, that the BBH merger rate is average within a Hubble time, and that single stars burst at $t=0$.
We can approximate the merger rate in generic star clusters to
\begin{equation}\label{eq1}
R_{\rm cl} = \frac{N_{\rm cl}N_{\rm 1g}f_{\rm BH}} {2\tau_{\rm Hub}},
\end{equation}
where $N_{\rm cl}$ is the number density of star clusters,
$N_{\rm 1g}$ (i.e., $N_{\rm BH}$) is the total number of first-generation (1g) BHs formed from the collapse of stars in the cluster,
and $f_{\rm BH}$ represents the fraction of BHs that form BH--BH binaries, harden, and merge within a Hubble time~$\tau_{\rm Hub}$.
For example, if the cluster has $N_{\rm 1g} = 100$  and $f_{\rm BH} = 1$, there are 50 mergers.
We note that $f_{\rm BH}$ in practice is not constant, which depends on mass ratio, total mass, and spins of BBHs.
We then relax the above assumption and consider hierarchical mergers, rewriting Eq.~(\ref{eq1}) as
\begin{equation}\label{eq2}
R_{\rm cl} \approx \frac{N_{\rm cl}} {\tau_{\rm Hub}} (\frac{1}{4}N_{\rm 1g}f_{\rm BH}+\frac{1}{4}N_{\rm 2g} f_{\rm BH}).
\end{equation}
Hierarchical mergers will occur in the 2g population with $N_{\rm 2g} = N_{\rm 1g}-N_{\rm 1g}f_{\rm BH}+N_{\rm rem}P_{\rm ret}$, with $N_{\rm rem} = R_{\rm cl}\tau_{\rm Hub} / N_{\rm cl}$ being the total BBH merger number and $P_{\rm ret}$ the retention probability of the remnant BHs in the cluster.
We expect that the relationship between the 1g and 2g populations is that $N_{\rm 2g} \sim N_{\rm 1g}$, because the number of BHs in the cluster that do not participate in mergers is much greater than the number that do, and some of the remnant BHs can be retained by the host cluster if the escape speed of the host is larger than their kick velocities.
This illustrates that Eq.~(\ref{eq2}) can be derived naturally and correctly.
For $N_{\rm 2g}$ (below), we use the exact value with $N_{\rm 1g}-N_{\rm 1g}f_{\rm BH}+N_{\rm rem}P_{\rm ret}$, and not the approximate value ($\sim N_{\rm 1g}$).
Moreover, the mergers of a higher generation than 2g + 2g  are extremely rare, and we therefore neglect them~\citep{Gerosa-Davide-PhysRevD.100.041301,Rodriguez-2019-Zevin-PhRvD.100d3027}.

The 2g population ($N_{\rm 2g}$) includes two parts: one is an initial BH population ($N_{\rm 1g}(1-f_{\rm BH})$) and the other is a remnant BH population ($N_{\rm rem}P_{\rm ret}$); here, we limit our discussion to the 2g population without reference to the 1g population.
Therefore, the merger from the 2g population could be a first-generation (1g + 1g) merger or a hierarchy (1g + 2g or 2g + 2g merger; we note that the merger from the 1g population must be a 1g + 1g merger).
A key piece of information is whether any of the BHs come from the remnant BH population, because the condition for a hierarchical merger to occur is that at least one of the BHs comes from the remnant of a previous merger (i.e., the BH in the remnant BH population).
Here, we introduce a hierarchical probability factor $\epsilon$ to represent the probability that a merger produced from the 2g population is a hierarchy, and correspondingly $1-\epsilon$ stands for the probability that this merger is a 1g + 1g merger.
For $\epsilon$, it can also be understood as the ratio of the number of 1g + 1g mergers from the 2g population to the total number of mergers, including 1g + 1g, 1g + 2g, and 2g + 2g mergers in the 2g population.
If all the BHs come from the initial BH population, then the number of all combinations of 1g + 1g mergers from random pairings is ${\rm C}^2_{N_{\rm 1g}\left(1-f_{\rm BH}\right)}$ , where ${\rm C}^m_n = \frac{n(n-1)(n-2) \cdots (n-m+1)}{m!}$ is the number of all such combinations for taking $m$ different elements from $n$ ($n\geq m$) different elements at a time regardless of their order.
The number of all combinations of 1g + 1g, 1g + 2g, and 2g + 2g mergers in the 2g population is ${\rm C}^2_{N_{\rm 2g}}$.
Hence, $\epsilon$ has the following form:
\begin{align}\label{eq5}
\epsilon &= 1-\frac{{\rm C}^2_{N_{\rm 1g}\left(1-f_{\rm BH}\right)}}{{\rm C}^2_{N_{\rm 2g}}}\nonumber\\
&= 1-\frac{N_{\rm 1g}\left(1-f_{\rm BH}\right) \left[N_{\rm 1g}\left(1-f_{\rm BH}\right)-1\right]}{N_{\rm 2g}(N_{\rm 2g}-1)}
\approx 1-\left[\frac{N_{\rm 1g}\left(1-f_{\rm BH}\right)}{N_{\rm 2g}}\right]^2, % '&'表示对点
\end{align}
because $N_{\rm 1g}\left(1-f_{\rm BH}\right) \gg 1$.

The number of mergers from the 2g population ($N_{\rm 2g}$) is $\frac{1}{4} N_{\rm 2g}f_{\rm BH}$.
Thus, the number of hierarchical mergers should be $\epsilon$ multiplied by $\frac{1}{4} N_{\rm 2g}f_{\rm BH}$ in the cluster.
In addition, the total number of BBH mergers that come from both the 1g population and the 2g population is $N_{\rm rem}$.
Consequently, the fraction of hierarchical mergers is obtained by dividing these two items:
\begin{align}\label{eq4}
f_{\rm hier} & = \frac{\epsilon N_{\rm 2g}f_{\rm BH}} {4N_{\rm rem}}.
\end{align}
Therefore, the rate of hierarchical merger should be
\begin{align}\label{eq6}
R_{\rm hier} &= R_{\rm cl}f_{\rm hier}\nonumber\\
&= \frac{1}{2} R_{\rm cl}f_{\rm BH}P_{\rm ret}
\left[1-\frac{f_{\rm BH}P_{\rm ret}}
{4-f_{\rm BH}\left(4-2P_{\rm ret}\right)}\right].
\end{align}
Eq.~(\ref{eq6}) can be approximated to (this is verified in Fig.~{\MyFigA})
\begin{equation}\label{eq7}
R_{\rm hier} \sim \frac{1}{2} R_{\rm cl}f_{\rm BH}P_{\rm ret}.
\end{equation}
Equation~(\ref{eq7}) suggests that the rate of hierarchical merger is approximately proportional to the retention probability.
Interestingly, the form of Eq.~(\ref{eq7}) is essentially the same as that of Eq.~(17) of~\cite{Kimball-2020-ApJ.900.177} who based their calculations on  a comparison with simulations~\citep{Rodriguez-2019-Zevin-PhRvD.100d3027}.
Then, combining Eqs.~(\ref{eq6}) and (\ref{eq7}), we can rewrite Eq.~(\ref{eq4}) into an approximate but simple form as
\begin{equation}\label{eq8}
f_{\rm hier} \sim \frac{1}{2}f_{\rm BH}P_{\rm ret}.
\end{equation}

We further consider the branching ratio ($\Gamma_{\rm \frac{1g + 2g}{2g + 2g}}$) of 1g + 2g to 2g + 2g merger rates.
In the same way as Eq.~(\ref{eq5}), we can obtain hierarchical probability factors $\epsilon_{\rm 1g+2g} = 1-\frac{{\rm C}^2_{N_{\rm 1g}\left(1-f_{\rm BH}\right)}}{{\rm C}^2_{N_{\rm 2g}}}-
\frac{{\rm C}^2_{N_{\rm rem}P_{\rm ret}}}{{\rm C}^2_{N_{\rm 2g}}}$ for 1g + 2g mergers and $\epsilon_{\rm 2g+2g} =\frac{{\rm C}^2_{N_{\rm rem}P_{\rm ret}}}{{\rm C}^2_{N_{\rm 2g}}}$ for 2g + 2g mergers.
Therefore, we expect
\begin{align}\label{eq9}
\Gamma_{\rm \frac{1g + 2g}{2g + 2g}} &= \frac{1-
\frac{{\rm C}^2_{N_{\rm 1g}\left(1-f_{\rm BH}\right)}}{{\rm C}^2_{N_{\rm 2g}}}-
\frac{{\rm C}^2_{N_{\rm rem}P_{\rm ret}}}{{\rm C}^2_{N_{\rm 2g}}}}
{\frac{{\rm C}^2_{N_{\rm rem}P_{\rm ret}}}{{\rm C}^2_{N_{\rm 2g}}}}\nonumber\\
& \sim \frac{4(1/f_{\rm BH}-1)}{P_{\rm ret}}.
\end{align}
If we assume that the fraction of BHs $f_{\rm BH} \lesssim 20\%$~\citep{Antonini-2014ApJ...794..106A,2018ApJ...866...66M} and the retention probability $P_{\rm ret} \lesssim 80\%$ (\cite{Mahapatra-Parthapratim-2021ApJ...918L..31M}, also see Sect.~\ref{ret}), the branching ratio of hierarchical mergers is $\gtrsim 20$.
This indicates that more than $\sim 95\%$ of hierarchical mergers in clusters are 1g + 1g mergers rather than 2g + 2g mergers.

It is noteworthy that
(1) some of the 2g binaries we calculate as successful mergers may not merge within a Hubble time because they must go through 1g mergers first;
(2) we add remnant BHs to the 2g population at once, but the process should take place during the whole Hubble time; and
(3) the densities of star clusters and the numbers of BHs in them  are varying over cosmic time, and it is therefore unrealistic to expect them to be well-approximated as constants over a Hubble time, although they may be constant over shorter evolutionary windows.
For these reasons, the simplicity of our assumptions should be taken into account when interpreting our results.
%
%%%%%%%%%%%%%%%%%%%%%%%%%%%%%%%%%%%%%%%%%%%%%%%%%%%%%%%%%%%%%%%%%%%%%%%%%%%%%%%
%---------------------------------------------------------------------------------

%
\begin{figure*}[t!]
\begin{center}
\includegraphics[width=1\hsize]{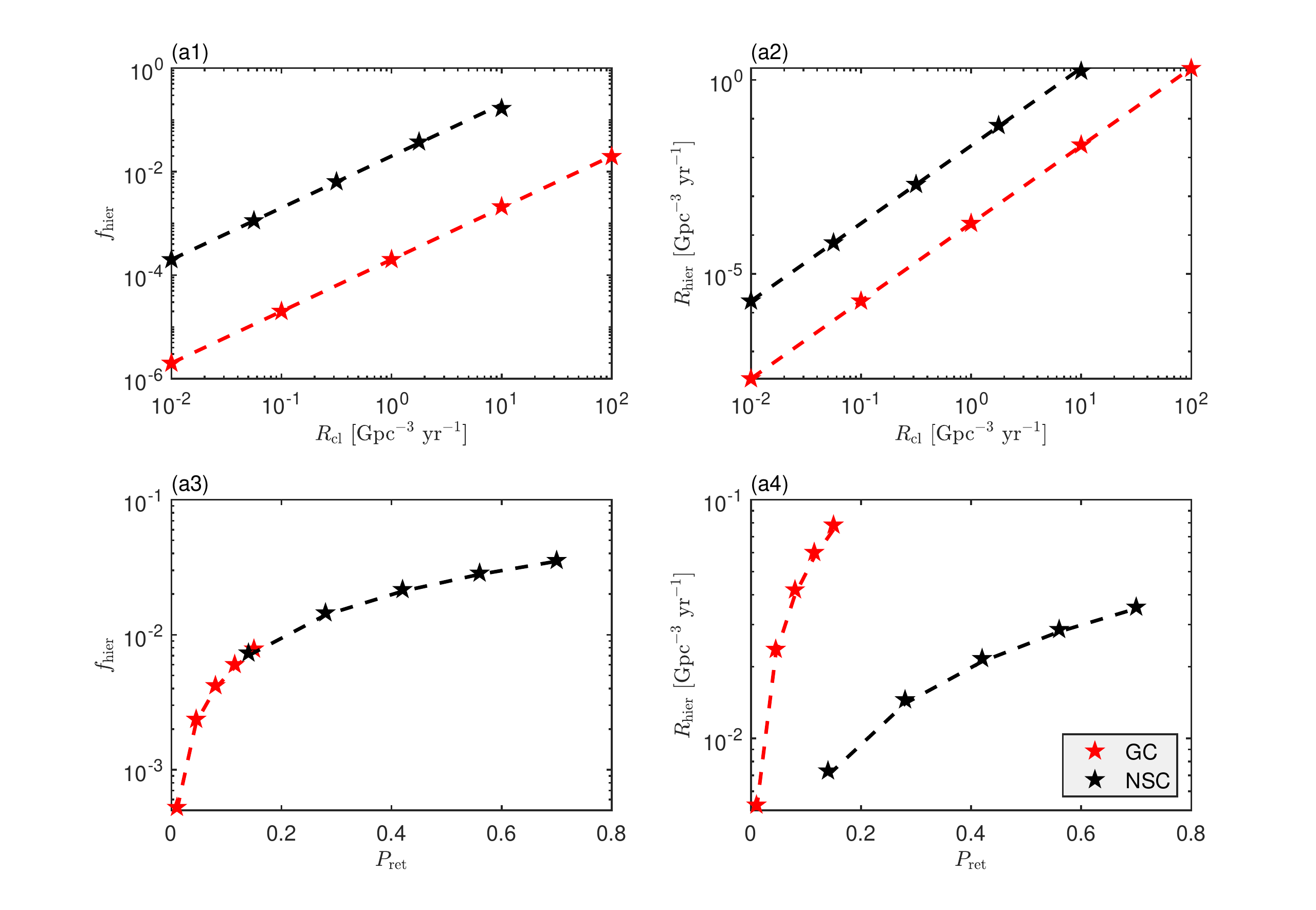}
\caption{Fractions (\emph{left}) and rates (\emph{right}) of hierarchical mergers as a function of the BBH merger rates (\emph{top}) and retention probabilities (\emph{bottom}) in GCs (red) and NSCs (black).
The derived values of the fractions and rates are obtained according to Eqs.~(\ref{eq4}) and (\ref{eq6}) (star symbols), respectively.
The approximate values were calculated using Eqs.~(\ref{eq8}) and (\ref{eq7}) (dashed line).
The fractions and rates are plotted as a function of BBH merger rates (retention probabilities) with fixed retention probabilities of $4\%$ and $40\%$ (BBH merger rates fixed to $10~\rate$ and $1~\rate$) for GCs and NSCs, respectively.
$N_{\rm cl} \sim 1~\rm Mpc^{-3}$ and $\sim 0.01~\rm Mpc^{-3}$  and $N_{\rm BH} \sim 2000$ and $\sim 2\times10^4$ are adopted for GCs and NSCs, respectively.}
\label{MyFigA}
\end{center}
\end{figure*}

\subsection{Density of star clusters and BH numbers}
In general, roughly one star per $1000~\Msun$ will be massive enough to form a BH, which is simply a result of the initial mass function of stars and the lowest birth mass needed to form a BH ($\sim 20~\Msun$)~\citep{2009MNRAS.395.2127O}.
Therefore, the number of BHs ---which is going to be a strong function of radius in star clusters (especially in NSCs)--- will scale with the mass of the cluster, and there is no unique answer because star clusters have different masses~\citep{2009MNRAS.395.2127O,2013ApJ...777..103T}.
We simply assume that there are $\sim 2000$ ($\sim 2\times10^4$) BHs in a GC (NSC) with a space density of around $1$ ($0.01$) per $\rm Mpc^3$ ~\citep{Baldry-2012MNRAS.421..621B,Antonini-2014ApJ...794..106A,Antonini-2016ApJ...831..187A}.
In comparison, there are $\sim 1{-}4\times10^4$ stellar mass BHs in the central parsec of the Milky Way~\citep{2000ApJ...545..847M,Antonini-2014ApJ...794..106A,10.1093/mnras/sty1262}.

\subsection{Retention probability of remnant BHs}\label{ret}
The retention probabilities of remnant BHs of $\sim 3\%$ and $46\%$ were obtained in GCs and NSCs respectively, from~\cite{Doctor-Zoheyr-2021ApJ...914L..18D} who used a NR surrogate waveform model~\citep{Varma-Vijay-PhysRevResearch.1.033015,Varma-Vijay-PhysRevLett.122.011101}.
Following the calculations of~\cite{Mahapatra-Parthapratim-2021ApJ...918L..31M}, we find that clusters with escape speeds of $50~\rm{km~s^{-1}}$ and $250~\rm{km~s^{-1}}$ can retain $\sim 1\%{-}12\%$ and $\sim 14\%{-}70\%$ of the GWTC-2 events, respectively.
Here, we adopt the retention probabilities of $4\%$ and $40\%$ in GCs and NSCs, respectively, as fiducial values and consider a wide range for comparison below.

\section{Rate of hierarchical mergers}\label{Rate}
We assume that the BBH merger rate is uniform over the $\sim 10~{\rm Gyr}$ history of the Universe, allowing us to obtain the fractions and rates of hierarchical mergers according to Eqs.~(\ref{eq4}) and (\ref{eq6}) (star symbols) respectively, which we show in Fig.~{\MyFigA}.
We also use simplified Eqs.~(\ref{eq7}) and (\ref{eq8}) (dashed line) to verify that they are correct.

From Fig.~{\MyFigA}~(a1), we see that the fraction of hierarchical mergers in GCs is two orders of magnitude smaller than that in NSCs if they have the same BBH merger rates.
If we consider that the typical BBH merger rates are $\sim 10~\rate$ and $\sim 1~\rate$ in GCs and NSCs~\citep{2022LRR....25....1M} respectively, then the fraction is $\sim 2\%$ in NSCs, while this value is $\sim 0.2\%$ in GCs.
We find that GCs and NSCs have hierarchical merger rates of $\sim 0.02~\rate$ (see Fig.~{\MyFigA}~(a2)).

Generally, the escape speeds in NSCs are larger than those in GCs, which causes the retention probability of remnant BHs in NSCs to be about one order of magnitude larger than that in GCs.
This therefore suggests that hierarchical mergers are rarer in GCs.
However, from Fig.~{\MyFigA}~(a2) we find that the rates of hierarchical mergers in GCs and NSCs are the same, considering that the BBH merger rate in GCs is ten times larger than that in NSCs~\citep{2022LRR....25....1M}.
We note that the similar rates of hierarchical mergers of GCs and NSCs are due to the fact that the retention fraction is taken here to be $4\%$ and $40\%$  (see Sect.~\ref{ret}), respectively, and under the assumption that mergers have no dependency on physical parameters such as masses and mass ratio (see Sect.~\ref{Sec:Method}).
Further, we calculate the ratio of hierarchical mergers in NSCs to those in GCs using Eq.~(\ref{eq7}):
\begin{align}\label{eq10}
\Gamma_{\rm \frac{NSC}{GC}}^{\rm (hier)} &= \frac{f_{\rm BH,NSC}}{f_{\rm BH,GC}}\frac{R_{\rm NSC}}{R_{\rm GC}}\frac{P_{\rm ret,NSC}}{P_{\rm ret,GC}},
\end{align}
where $f_{\rm BH,NSC}$ ($f_{\rm BH,GC}$) is the fraction of BHs that form BH--BH binaries, harden, and merge in NSCs (GCs), and
$P_{\rm ret,NSC}$ and $P_{\rm ret,GC}$ are the retention probabilities for NSCs and GCs, respectively.
.
We set that $R_{\rm NSC}/R_{\rm GC} \sim 0.1$~\citep{2022LRR....25....1M} and $P_{\rm ret,NSC}/P_{\rm ret,GC} \sim 10$~\citep{Doctor-Zoheyr-2021ApJ...914L..18D,Mahapatra-Parthapratim-2021ApJ...918L..31M}.
Therefore, we obtain a ratio of  hierarchical merger rate in NSCs to that in GCs of $\sim 1$ if $f_{\rm BH,NSC} \approx f_{\rm BH,GC}$.
Even if the ratio ($f_{\rm BH,NSC}/f_{\rm BH,GC}$) is not equal to 1, its fluctuation would not be significantly large.
In consequence, the hierarchical merger rate in GCs is comparable to that in NSCs.
\begin{table*}
\begin{center}
\caption{Fractions, rates, and branching ratios of hierarchical mergers for different cases.}\label{MyTabA}
\centering{
\resizebox{\textwidth}{18mm}{
\begin{tabular}{lcccccccccc}
\hline\hline\rule{0pt}{12pt}
Parameter &\multicolumn{4}{c}{GC (1 $\rm Mpc^{-3}$)} &\multicolumn{6}{c}{NSC (0.01 $\rm Mpc^{-3}$)}\\\hline\rule{0pt}{10pt}
BH number ($N_{\rm BH}$) &$2,000$ &$2,000$ &$2,000$ &$4,000$ &$1,000$&$1\times10^4$ &$2\times10^4$ &$2\times10^4$ &$5\times10^4$ & $10^5$\\\rule{0pt}{10pt}
BBH merger rate ($R_{\rm cl}~[\rate]$) &4 &10 &10 &40 &0.01 &0.4 &1 &2 &4 &10\\\rule{0pt}{10pt}
Retention probability ($P_{\rm ret}$) &0.03 &0.04 &0.1 &0.1 &0.2&0.3 &0.4 &0.6 &0.4 &0.7\\\hline\rule{0pt}{10pt}
Fraction of hierarchical mergers ($f_{\rm hier}$) &$6\times10^{-4}$ &$2\times10^{-3}$ &$5\times10^{-3}$ &0.01 &$2\times10^{-3}$ &0.01 &0.02 &0.06 &0.03 &0.07\\\rule{0pt}{10pt}
Hierarchical merger rate ($R_{\rm hier}~[\rate]$)& $2.4\times10^{-3}$ &0.02 &0.05 &0.4 &$2\times10^{-5}$ &$4.8\times10^{-3}$ &0.02 &0.12 &0.13 &0.7\\\rule{0pt}{10pt}
Branching ratio ($\Gamma_{\rm \frac{1g + 2g}{2g + 2g}}$) &3200 &900 &360 &160 &980 &153 &90 &26.7 &52.5 &22.9
\\\hline
%\bottomrule[0.6pt]
\end{tabular}}}
\end{center}
\end{table*}

We relax the retention probability for GCs up to $\sim 15\%$ in Fig.~{\MyFigA}~(a3) and see that the fraction of hierarchical mergers in GCs can reach as high as that in NSCs.
The corresponding hierarchical merger rate in GCs can even exceed that in NSCs (see Fig.~{\MyFigA}~(a4)).
We also adopt wide ranges in BH numbers and consider different BBH merger rates and retention probabilities to estimate fractions, rates, and branching ratios.
The obtained results are listed in Table~{\MyTabA}.
We see that the hierarchical merger rate is affected by significant uncertainty, with a wide range from ${\mathcal O(10^{-5})}$ to ${\mathcal O(1)}~\rate$.
Overall, the rate of hierarchical mergers is very small.
If we assume that all the BBH mergers detected by LIGO-Virgo came from GCs or NSCs, then at most one event might instead be a hierarchical merger.
We note that a wide range of BBH merger rates is adopted in Table~{\MyTabA}.
Therefore, it is normal that merger rates may be overestimated or underestimated compared with the results from the LVK Collaboration, who found the BBH merger rate to be between $17.9~\rate$ and $44~\rate$ at a fiducial redshift ($z = 0.2$)~\citep{LVK-arXiv-GWTC-3-population-211103634}.

\section{Discussion}\label{Discussion}
~\cite{Rodriguez-2019-Zevin-PhRvD.100d3027} used a grid of 96 dynamical models of dense star clusters and a cosmological model of cluster formation to explore the production of hierarchical BBH mergers.
These latter authors found that if all initial BHs that formed from the collapse of stars are nonspinning (i.e., BH dimensionless spin parameter $\chi = 0$), then more than $10\%$ of mergers are hierarchical mergers.
Following the method of~\cite{DG-PhysRevD.104.084002} to pair binaries, we estimate the retention probability of 1g + 1g mergers by calculating kick velocities of merging binaries using NR fitting formulas~\citep{Gonzalez-Jose-2007ApJ...659L...5C}.
We find that more than $80\%$ of remnant BHs would be retained by clusters with escape speeds of $\sim 200~\rm{km~s^{-1}}$ if all 1g BHs are nonspinning.
This is relatively high compared to existing observations and calculations~\citep{Fragione-2021-Loeb-MNRAS.502.3879,Mahapatra-Parthapratim-2021ApJ...918L..31M,
Kimball-2021-ApJ.915L35}, therefore the contribution of hierarchical mergers from GCs to the BBH merger rate might drop from $\sim 10\%$ to $\lesssim 1\%$ for moderate $\chi \sim 0.5$~\citep{Rodriguez-2019-Zevin-PhRvD.100d3027}.
This is consistent with our result  that the fractions of hierarchical mergers $f_{\rm hier}$ are all less than 10\%; see Table~{\MyTabA} (also see Fig.~{\MyFigA}~(a1) and (a3)).
The simplified model of~\cite{Gerosa-Davide-PhysRevD.100.041301} also shows a relatively small fraction of hierarchical mergers in clusters if the BH spin is as much as 0.2.
Moreover,~\cite{2021Symm...13.1678M} found that hierarchical BBHs in NSCs account for $\sim 10^{-2}{-}0.2~\rate$, which broadly agrees with our estimates (see Fig.~{\MyFigA}~(a4)).

The 2g BHs may preferentially form binaries in dynamical interactions due to more massive components.
As a result, the fraction $f_{\rm BH}$ is different in the 1g population and the 2g population in Eq.~(\ref{eq2}).
Furthermore, there are some initial binaries in the 1g population, which would make $f_{\rm BH}$ smaller in 1g + 1g mergers and larger in higher generation mergers.
To justify the use of the same $f_{\rm BH}$ for the 1g and 2g populations, we set the fraction in the 2g population to $n$ multiplied by $f_{\rm BH}$ in the 1g population.
The factor $n$ is larger than 1 but limited to $1/f_{\rm BH}$.
Then, Eq.~(\ref{eq2}) is
\begin{equation}\label{eq10}
R_{\rm cl} \approx \frac{N_{\rm cl}} {\tau_{\rm Hub}} (\frac{1}{4}N_{\rm 1g}f_{\rm BH}+\frac{1}{4}N_{\rm 2g} nf_{\rm BH}).
\end{equation}
We obtain the rate of hierarchical mergers using Eqs.~(\ref{eq5}) and (\ref{eq4}):
\begin{align}\label{eq11}
R^{(n)}_{\rm hier} = \frac{1}{2} nR_{\rm cl}f_{\rm BH}P_{\rm ret}
\left[1-\frac{f_{\rm BH}P_{\rm ret}}
{4-f_{\rm BH}\left(4-2P_{\rm ret}\right)}\right].
\end{align}
We can define a ratio $\Gamma_{\frac{n}{1}}$, which is equal to the rate from Eq.~(\ref{eq11}) divided by that from Eq.~(\ref{eq6}):
\begin{equation}\label{eq10}
\Gamma_{\frac{n}{1}} = \frac{R^{(n)}_{\rm hier}}{R_{\rm hier}}=n.% 1 \leq n \leq  \frac{1}{f_{\rm BH}}.
\end{equation}
We expect $n$ to be between 1 and 2 in general due to the fact that the binary needs to survive few-body interactions, and therefore the difference between considering the change of the fraction and not considering the change of the fraction results in a change of less than a factor two for the rates of hierarchical mergers.
This effect is relatively small and the results from Eq.~(\ref{eq6}) are reliable.
%
%%%%%%%%%%%%%%%%%%%%%%%%%%%%%%%%%%%%%%%%%%%%%%%%%%%%%%%%%%%%%%%%%%%%%%%%%%%%%%%
%---------------------------------------------------------------------------------
%
\section{Conclusion}\label{Conclusion}
The kick can be extracted from the GW signals.
This information can then be used to obtain the retention probability of remnant BHs in different types of star clusters assuming all GWTC events occurred in clusters.

In this study, we attempted to estimate the rates of hierarchical merger from the retention probability of remnant BHs.
A simple formula is given by Eq.~(\ref{eq7}) that suggests the rate of hierarchical merger is proportional to the retention probability.  We find that
$\sim 2\%$ of BBH mergers in NSCs may be hierarchical mergers, while this percentage in GCs is a few tenths of a percent.
However, this does not mean the rate of hierarchical merger in GCs is smaller than that in NSCs.
We find that hierarchical merger rates in GCs and NSCs are about the same with $\sim {\mathcal O(10^{-2})}~{\rm Gpc^{-3}~yr^{-1}}$ because the BBH merger rate in GCs is actually higher than that in NSCs.

Similar conclusions on hierarchical mergers (and rates) have already been obtained by various
authors~\citep{Rodriguez-2019-Zevin-PhRvD.100d3027,
Kimball-2020-ApJ.900.177,2020ApJ...894..133A,2020MNRAS.498.4591F,2021MNRAS.505..339M,2021Symm...13.1678M} using even more sophisticated techniques (i.e., less approximate).
Nevertheless, simple analytic arguments can provide useful insights because our results are not based on simulations.
Moreover, we also point out that although the efficiency of hierarchical mergers in GCs is lower than that in NSCs, this does not mean that the hierarchical merger rate in GCs is lower than that in NSCs (see Sect.~\ref{Rate}).
This provides the warning that if an event detected by LIGO-Virgo is determined as a hierarchical merger, then the probability that it originates from a GC is approximately equal to the probability that it originates from an NSC.

Understanding the distribution of BHs at the centers of galaxies is crucial for making predictions about the expected merger rates.
We expect that with third-generation GW detectors in operation~\citep{Einstein-Telescope-2010CQGra..27s4002P,third-generation-2010CQGra..27h4007P,Abbott-2017CQGra..34d4001A}, the increasing data on GW events will help us to constrain the retention probability and merger rate more precisely.
%
%%%%%%%%%%%%%%%%%%%%%%%%%%%%%%%%%%%%%%%%%%%%%%%%%%%%%%%%%%%%%%%%%%%%%%%%%%%%%%%
%---------------------------------------------------------------------------------
\begin{acknowledgements}
G.-P.~Li is grateful to Da-Bin Lin, Ilya Mandel, Xiao-Yan Li, Jia-Xin Cao, Cheng Gao, and Hao-Qiang Zhang for useful discussions.
This work is supported by the National Natural Science Foundation of China (Grant No. 11773007) and the Guangxi Science Foundation (Grant No. 2018GXNSFFA281010).
\end{acknowledgements}
%%%%%%%%%%%%%%%%%%%%%%%%%%%%%%%%%%%%%%%%%%%%%%%%%%%%%%%%%%%%%%%%%%%%%%%%%%%%
%\bibliographystyle{aa}
%\bibliography{hier}

\begin{thebibliography}{56}
\expandafter\ifx\csname natexlab\endcsname\relax\def\natexlab#1{#1}\fi

\bibitem[{{Abbott} {et~al.}(2016){Abbott}, {Abbott}, {Abbott}, {Abernathy},
  {Acernese}, {Ackley}, {Adams}, {Adams}, {Addesso}, {Adhikari}, \&
  et~al.}]{GW150914}
{Abbott}, B.~P., {Abbott}, R., {Abbott}, T.~D., {et~al.} 2016, \prl, 116,
  061102

\bibitem[{{Abbott} {et~al.}(2017){Abbott}, {Abbott}, {Abbott}, {Abernathy},
  {Ackley}, {Adams}, {Addesso}, {Adhikari}, {Adya}, {Affeldt}, \&
  et~al.}]{Abbott-2017CQGra..34d4001A}
{Abbott}, B.~P., {Abbott}, R., {Abbott}, T.~D., {et~al.} 2017, Classical and
  Quantum Gravity, 34, 044001

\bibitem[{{Abbott} {et~al.}(2019){Abbott}, {Abbott}, {Abbott}, {Abraham},
  {Acernese}, {Ackley}, {Adams}, {Adhikari}, {Adya}, {Affeldt}, \&
  et~al.}]{GWTC-1}
{Abbott}, B.~P., {Abbott}, R., {Abbott}, T.~D., {et~al.} 2019, Physical Review
  X, 9, 031040

\bibitem[{{Abbott} {et~al.}(2021){Abbott}, {Abbott}, {Abraham}, {Acernese},
  {Ackley}, {Adams}, {Adams}, {Adhikari}, {Adya}, {Affeldt}, \&
  et~al.}]{GWTC-2}
{Abbott}, R., {Abbott}, T.~D., {Abraham}, S., {et~al.} 2021, Physical Review X,
  11, 021053

\bibitem[{{Abbott} {et~al.}(2020{\natexlab{a}}){Abbott}, {Abbott}, {Abraham},
  {Acernese}, {Ackley}, {Adams}, {Adhikari}, {Adya}, {Affeldt}, {Agathos}, \&
  et~al.}]{GW190412}
{Abbott}, R., {Abbott}, T.~D., {Abraham}, S., {et~al.} 2020{\natexlab{a}},
  \prd, 102, 043015

\bibitem[{{Abbott} {et~al.}(2020{\natexlab{b}}){Abbott}, {Abbott}, {Abraham},
  {Acernese}, {Ackley}, {Adams}, {Adhikari}, {Adya}, {Affeldt}, {Agathos}, \&
  et~al.}]{GW190521}
{Abbott}, R., {Abbott}, T.~D., {Abraham}, S., {et~al.} 2020{\natexlab{b}},
  \prl, 125, 101102

\bibitem[{{Acernese} {et~al.}(2015){Acernese}, {Agathos}, {Agatsuma}, {Aisa},
  {Allemandou}, {Allocca}, {Amarni}, {Astone}, {Balestri}, {Ballardin}, \&
  et~al.}]{Virgo-2015}
{Acernese}, F., {Agathos}, M., {Agatsuma}, K., {et~al.} 2015, Classical and
  Quantum Gravity, 32, 024001

\bibitem[{{Anagnostou} {et~al.}(2020){Anagnostou}, {Trenti}, \&
  {Melatos}}]{Anagnostou-2020arXiv201006161A}
{Anagnostou}, O., {Trenti}, M., \& {Melatos}, A. 2020, arXiv e-prints,
  arXiv:2010.06161

\bibitem[{{Antonini}(2014)}]{Antonini-2014ApJ...794..106A}
{Antonini}, F. 2014, \apj, 794, 106

\bibitem[{{Antonini} \& {Rasio}(2016)}]{Antonini-2016ApJ...831..187A}
{Antonini}, F. \& {Rasio}, F.~A. 2016, \apj, 831, 187

\bibitem[{{Arca Sedda} {et~al.}(2020){Arca Sedda}, {Mapelli}, {Spera},
  {Benacquista}, \& {Giacobbo}}]{2020ApJ...894..133A}
{Arca Sedda}, M., {Mapelli}, M., {Spera}, M., {Benacquista}, M., \& {Giacobbo},
  N. 2020, \apj, 894, 133

\bibitem[{Baibhav {et~al.}(2021)Baibhav, Berti, Gerosa, Mould, \&
  Wong}]{DG-PhysRevD.104.084002}
Baibhav, V., Berti, E., Gerosa, D., Mould, M., \& Wong, K. W.~K. 2021, Phys.
  Rev. D, 104, 084002

\bibitem[{{Baldry} {et~al.}(2012){Baldry}, {Driver}, {Loveday}, {Taylor},
  {Kelvin}, {Liske}, {Norberg}, {Robotham}, {Brough}, {Hopkins}, {Bamford},
  {Peacock}, {Bland-Hawthorn}, {Conselice}, {Croom}, {Jones}, {Parkinson},
  {Popescu}, {Prescott}, {Sharp}, \& {Tuffs}}]{Baldry-2012MNRAS.421..621B}
{Baldry}, I.~K., {Driver}, S.~P., {Loveday}, J., {et~al.} 2012, \mnras, 421,
  621

\bibitem[{Calder\'on~Bustillo {et~al.}(2018)Calder\'on~Bustillo, Clark, Laguna,
  \& Shoemaker}]{Calderon-Bustillo-PhysRevLett.121.191102}
Calder\'on~Bustillo, J., Clark, J.~A., Laguna, P., \& Shoemaker, D. 2018, Phys.
  Rev. Lett., 121, 191102

\bibitem[{{Campanelli} {et~al.}(2007){Campanelli}, {Lousto}, {Zlochower}, \&
  {Merritt}}]{Gonzalez-Jose-2007ApJ...659L...5C}
{Campanelli}, M., {Lousto}, C., {Zlochower}, Y., \& {Merritt}, D. 2007, \apjl,
  659, L5

\bibitem[{{Dall'Amico}(2021)}]{Dall'Amico2021arXiv211202020D}
{Dall'Amico}, M. 2021, arXiv e-prints, arXiv:2112.02020

\bibitem[{{Dall'Amico} {et~al.}(2021){Dall'Amico}, {Mapelli}, {Di Carlo},
  {Bouffanais}, {Rastello}, {Santoliquido}, {Ballone}, \& {Arca
  Sedda}}]{Dall'Amico-2021MNRAS.508.3045D}
{Dall'Amico}, M., {Mapelli}, M., {Di Carlo}, U.~N., {et~al.} 2021, \mnras, 508,
  3045

\bibitem[{{Doctor} {et~al.}(2021){Doctor}, {Farr}, \&
  {Holz}}]{Doctor-Zoheyr-2021ApJ...914L..18D}
{Doctor}, Z., {Farr}, B., \& {Holz}, D.~E. 2021, \apjl, 914, L18

\bibitem[{{Fragione} {et~al.}(2022){Fragione}, {Kocsis}, {Rasio}, \&
  {Silk}}]{2022ApJ...927..231F}
{Fragione}, G., {Kocsis}, B., {Rasio}, F.~A., \& {Silk}, J. 2022, \apj, 927,
  231

\bibitem[{{Fragione} \& {Loeb}(2021)}]{Fragione-2021-Loeb-MNRAS.502.3879}
{Fragione}, G. \& {Loeb}, A. 2021, \mnras, 502, 3879

\bibitem[{{Fragione} \& {Silk}(2020)}]{2020MNRAS.498.4591F}
{Fragione}, G. \& {Silk}, J. 2020, \mnras, 498, 4591

\bibitem[{Generozov {et~al.}(2018)Generozov, Stone, Metzger, \&
  Ostriker}]{10.1093/mnras/sty1262}
Generozov, A., Stone, N.~C., Metzger, B.~D., \& Ostriker, J.~P. 2018, \mnras,
  478, 4030

\bibitem[{Gerosa \& Berti(2019)}]{Gerosa-Davide-PhysRevD.100.041301}
Gerosa, D. \& Berti, E. 2019, Phys. Rev. D, 100, 041301

\bibitem[{{Gerosa} \& {Fishbach}(2021)}]{Gerosa-Davide-2021-review}
{Gerosa}, D. \& {Fishbach}, M. 2021, Nature Astronomy, 5, 749

\bibitem[{{Hamers} \&
  {Safarzadeh}(2020)}]{Hamers_A-2020-Safarzadeh_M-ApJ.898.99}
{Hamers}, A.~S. \& {Safarzadeh}, M. 2020, \apj, 898, 99

\bibitem[{{Kimball} {et~al.}(2020){Kimball}, {Talbot}, {Berry}, {Carney},
  {Zevin}, {Thrane}, \& {Kalogera}}]{Kimball-2020-ApJ.900.177}
{Kimball}, C., {Talbot}, C., {Berry}, C. P.~L., {et~al.} 2020, \apj, 900, 177

\bibitem[{{Kimball} {et~al.}(2021){Kimball}, {Talbot}, {Berry}, {Zevin},
  {Thrane}, {Kalogera}, {Buscicchio}, {Carney}, {Dent}, {Middleton}, {Payne},
  {Veitch}, \& {Williams}}]{Kimball-2021-ApJ.915L35}
{Kimball}, C., {Talbot}, C., {Berry}, C. P.~L., {et~al.} 2021, \apjl, 915, L35

\bibitem[{{Li}(2022)}]{Li-PRD-2022-105.063006}
{Li}, G.-P. 2022, \prd, 105, 063006

\bibitem[{{Liu} \& {Lai}(2021)}]{Lai-2021MNRAS.502.2049L}
{Liu}, B. \& {Lai}, D. 2021, \mnras, 502, 2049

\bibitem[{{Mahapatra} {et~al.}(2021){Mahapatra}, {Gupta}, {Favata}, {Arun}, \&
  {Sathyaprakash}}]{Mahapatra-Parthapratim-2021ApJ...918L..31M}
{Mahapatra}, P., {Gupta}, A., {Favata}, M., {Arun}, K.~G., \& {Sathyaprakash},
  B.~S. 2021, \apjl, 918, L31

\bibitem[{{Mandel} \& {Broekgaarden}(2022)}]{2022LRR....25....1M}
{Mandel}, I. \& {Broekgaarden}, F.~S. 2022, Living Reviews in Relativity, 25, 1

\bibitem[{{Mandel} \& {Farmer}(2022)}]{Ilya-2022PhR...955....1M}
{Mandel}, I. \& {Farmer}, A. 2022, \physrep, 955, 1

\bibitem[{{Mapelli}(2021)}]{Mapelli-2021hgwa.bookE...4M}
{Mapelli}, M. 2021, in Handbook of Gravitational Wave Astronomy, 4

\bibitem[{{Mapelli} {et~al.}(2021{\natexlab{a}}){Mapelli}, {Dall'Amico},
  {Bouffanais}, {Giacobbo}, {Arca Sedda}, {Artale}, {Ballone}, {Di Carlo},
  {Iorio}, {Santoliquido}, \& {Torniamenti}}]{2021MNRAS.505..339M}
{Mapelli}, M., {Dall'Amico}, M., {Bouffanais}, Y., {et~al.} 2021{\natexlab{a}},
  \mnras, 505, 339

\bibitem[{{Mapelli} {et~al.}(2021{\natexlab{b}}){Mapelli}, {Santoliquido},
  {Bouffanais}, {Arca Sedda}, {Artale}, \& {Ballone}}]{2021Symm...13.1678M}
{Mapelli}, M., {Santoliquido}, F., {Bouffanais}, Y., {et~al.}
  2021{\natexlab{b}}, Symmetry, 13, 1678

\bibitem[{{McKernan} {et~al.}(2018){McKernan}, {Ford}, {Bellovary}, {Leigh},
  {Haiman}, {Kocsis}, {Lyra}, {Mac Low}, {Metzger}, {O'Dowd}, {Endlich}, \&
  {Rosen}}]{2018ApJ...866...66M}
{McKernan}, B., {Ford}, K.~E.~S., {Bellovary}, J., {et~al.} 2018, \apj, 866, 66

\bibitem[{{Michaely} \& {Naoz}(2022)}]{2022arXiv220515040M}
{Michaely}, E. \& {Naoz}, S. 2022, arXiv e-prints, arXiv:2205.15040

\bibitem[{{Michaely} \& {Perets}(2019)}]{2019ApJ...887L..36M}
{Michaely}, E. \& {Perets}, H.~B. 2019, \apjl, 887, L36

\bibitem[{{Michaely} \& {Perets}(2020)}]{2020MNRAS.498.4924M}
{Michaely}, E. \& {Perets}, H.~B. 2020, \mnras, 498, 4924

\bibitem[{{Miller} \& {Colbert}(2004)}]{Miller-2004IJMPD.13.1}
{Miller}, M.~C. \& {Colbert}, E.~J.~M. 2004, International Journal of Modern
  Physics D, 13, 1

\bibitem[{{Miralda-Escud{\'e}} \& {Gould}(2000)}]{2000ApJ...545..847M}
{Miralda-Escud{\'e}}, J. \& {Gould}, A. 2000, \apj, 545, 847

\bibitem[{{O'Leary} {et~al.}(2009){O'Leary}, {Kocsis}, \&
  {Loeb}}]{2009MNRAS.395.2127O}
{O'Leary}, R.~M., {Kocsis}, B., \& {Loeb}, A. 2009, \mnras, 395, 2127

\bibitem[{{Punturo} {et~al.}(2010{\natexlab{a}}){Punturo}, {Abernathy},
  {Acernese}, {Allen}, {Andersson}, {Arun}, {Barone}, {Barr}, {Barsuglia},
  {Beker}, {Beveridge}, {Birindelli}, {Bose}, {Bosi}, {Braccini}, {Bradaschia},
  {Bulik}, {Calloni}, {Cella}, {Chassande Mottin}, {Chelkowski}, {Chincarini},
  {Clark}, {Coccia}, {Colacino}, {Colas}, {Cumming}, {Cunningham}, {Cuoco},
  {Danilishin}, {Danzmann}, {De Luca}, {De Salvo}, {Dent}, {De Rosa}, {Di
  Fiore}, {Di Virgilio}, {Doets}, {Fafone}, {Falferi}, {Flaminio}, {Franc},
  {Frasconi}, {Freise}, {Fulda}, {Gair}, {Gemme}, {Gennai}, {Giazotto},
  {Glampedakis}, {Granata}, {Grote}, {Guidi}, {Hammond}, {Hannam}, {Harms},
  {Heinert}, {Hendry}, {Heng}, {Hennes}, {Hild}, {Hough}, {Husa}, {Huttner},
  {Jones}, {Khalili}, {Kokeyama}, {Kokkotas}, {Krishnan}, {Lorenzini},
  {L{\"u}ck}, {Majorana}, {Mandel}, {Mandic}, {Martin}, {Michel}, {Minenkov},
  {Morgado}, {Mosca}, {Mours}, {M{\"u}ller{\textendash}Ebhardt}, {Murray},
  {Nawrodt}, {Nelson}, {Oshaughnessy}, {Ott}, {Palomba}, {Paoli}, {Parguez},
  {Pasqualetti}, {Passaquieti}, {Passuello}, {Pinard}, {Poggiani}, {Popolizio},
  {Prato}, {Puppo}, {Rabeling}, {Rapagnani}, {Read}, {Regimbau}, {Rehbein},
  {Reid}, {Rezzolla}, {Ricci}, {Richard}, {Rocchi}, {Rowan}, {R{\"u}diger},
  {Sassolas}, {Sathyaprakash}, {Schnabel}, {Schwarz}, {Seidel}, {Sintes},
  {Somiya}, {Speirits}, {Strain}, {Strigin}, {Sutton}, {Tarabrin},
  {Th{\"u}ring}, {van den Brand}, {van Leewen}, {van Veggel}, {van den Broeck},
  {Vecchio}, {Veitch}, {Vetrano}, {Vicere}, {Vyatchanin}, {Willke}, {Woan},
  {Wolfango}, \& {Yamamoto}}]{Einstein-Telescope-2010CQGra..27s4002P}
{Punturo}, M., {Abernathy}, M., {Acernese}, F., {et~al.} 2010{\natexlab{a}},
  Classical and Quantum Gravity, 27, 194002

\bibitem[{{Punturo} {et~al.}(2010{\natexlab{b}}){Punturo}, {Abernathy},
  {Acernese}, {Allen}, {Andersson}, {Arun}, {Barone}, {Barr}, {Barsuglia},
  {Beker}, {Beveridge}, {Birindelli}, {Bose}, {Bosi}, {Braccini}, {Bradaschia},
  {Bulik}, {Calloni}, {Cella}, {Chassande Mottin}, {Chelkowski}, {Chincarini},
  {Clark}, {Coccia}, {Colacino}, {Colas}, {Cumming}, {Cunningham}, {Cuoco},
  {Danilishin}, {Danzmann}, {De Luca}, {De Salvo}, {Dent}, {Derosa}, {Di
  Fiore}, {Di Virgilio}, {Doets}, {Fafone}, {Falferi}, {Flaminio}, {Franc},
  {Frasconi}, {Freise}, {Fulda}, {Gair}, {Gemme}, {Gennai}, {Giazotto},
  {Glampedakis}, {Granata}, {Grote}, {Guidi}, {Hammond}, {Hannam}, {Harms},
  {Heinert}, {Hendry}, {Heng}, {Hennes}, {Hild}, {Hough}, {Husa}, {Huttner},
  {Jones}, {Khalili}, {Kokeyama}, {Kokkotas}, {Krishnan}, {Lorenzini},
  {L{\"u}ck}, {Majorana}, {Mandel}, {Mandic}, {Martin}, {Michel}, {Minenkov},
  {Morgado}, {Mosca}, {Mours}, {M{\"u}ller-Ebhardt}, {Murray}, {Nawrodt},
  {Nelson}, {Oshaughnessy}, {Ott}, {Palomba}, {Paoli}, {Parguez},
  {Pasqualetti}, {Passaquieti}, {Passuello}, {Pinard}, {Poggiani}, {Popolizio},
  {Prato}, {Puppo}, {Rabeling}, {Rapagnani}, {Read}, {Regimbau}, {Rehbein},
  {Reid}, {Rezzolla}, {Ricci}, {Richard}, {Rocchi}, {Rowan}, {R{\"u}diger},
  {Sassolas}, {Sathyaprakash}, {Schnabel}, {Schwarz}, {Seidel}, {Sintes},
  {Somiya}, {Speirits}, {Strain}, {Strigin}, {Sutton}, {Tarabrin}, {van den
  Brand}, {van Leewen}, {van Veggel}, {van den Broeck}, {Vecchio}, {Veitch},
  {Vetrano}, {Vicere}, {Vyatchanin}, {Willke}, {Woan}, {Wolfango}, \&
  {Yamamoto}}]{third-generation-2010CQGra..27h4007P}
{Punturo}, M., {Abernathy}, M., {Acernese}, F., {et~al.} 2010{\natexlab{b}},
  Classical and Quantum Gravity, 27, 084007

\bibitem[{{Raveh} {et~al.}(2022){Raveh}, {Michaely}, \&
  {Perets}}]{2022MNRAS.514.4246R}
{Raveh}, Y., {Michaely}, E., \& {Perets}, H.~B. 2022, \mnras, 514, 4246

\bibitem[{{Rodriguez} {et~al.}(2019){Rodriguez}, {Zevin}, {Amaro-Seoane},
  {Chatterjee}, {Kremer}, {Rasio}, \&
  {Ye}}]{Rodriguez-2019-Zevin-PhRvD.100d3027}
{Rodriguez}, C.~L., {Zevin}, M., {Amaro-Seoane}, P., {et~al.} 2019, \prd, 100,
  043027

\bibitem[{{The LIGO Scientific Collaboration} {et~al.}(2015){The LIGO
  Scientific Collaboration}, {Aasi}, {Abbott}, {Abbott}, {Abbott}, {Abernathy},
  {Ackley}, {Adams}, {Adams}, {Addesso}, \& et~al.}]{LIGO-2015}
{The LIGO Scientific Collaboration}, {Aasi}, J., {Abbott}, B.~P., {et~al.}
  2015, Classical and Quantum Gravity, 32, 074001

\bibitem[{{The LIGO Scientific Collaboration} {et~al.}(2021{\natexlab{a}}){The
  LIGO Scientific Collaboration}, {the Virgo Collaboration}, {Abbott},
  {Abbott}, {Acernese}, {Ackley}, {Adams}, {Adhikari}, {Adhikari}, {Adya}, \&
  et~al.}]{GWTC-2.1}
{The LIGO Scientific Collaboration}, {the Virgo Collaboration}, {Abbott}, R.,
  {et~al.} 2021{\natexlab{a}}, arXiv e-prints, arXiv:2108.01045

\bibitem[{{The LIGO Scientific Collaboration} {et~al.}(2021{\natexlab{b}}){The
  LIGO Scientific Collaboration}, {the Virgo Collaboration}, {the KAGRA
  Collaboration}, {Abbott}, {Abbott}, {Acernese}, {Ackley}, {Adams},
  {Adhikari}, {Adhikari}, \& et~al.}]{GWTC-3}
{The LIGO Scientific Collaboration}, {the Virgo Collaboration}, {the KAGRA
  Collaboration}, {et~al.} 2021{\natexlab{b}}, arXiv e-prints, arXiv:2111.03606

\bibitem[{{The LIGO Scientific Collaboration} {et~al.}(2021{\natexlab{c}}){The
  LIGO Scientific Collaboration}, {the Virgo Collaboration}, {the KAGRA
  Collaboration}, {Abbott}, {Abbott}, {Acernese}, {Ackley}, {Adams},
  {Adhikari}, {Adhikari}, \& et~al.}]{LVK-arXiv-GWTC-3-population-211103634}
{The LIGO Scientific Collaboration}, {the Virgo Collaboration}, {the KAGRA
  Collaboration}, {et~al.} 2021{\natexlab{c}}, arXiv e-prints, arXiv:2111.03634

\bibitem[{{Tsang}(2013)}]{2013ApJ...777..103T}
{Tsang}, D. 2013, \apj, 777, 103

\bibitem[{{Varma} {et~al.}(2022){Varma}, {Biscoveanu}, {Islam}, {Shaik},
  {Haster}, {Isi}, {Farr}, {Field}, \& {Vitale}}]{2022PhRvL.128s1102V}
{Varma}, V., {Biscoveanu}, S., {Islam}, T., {et~al.} 2022, \prl, 128, 191102

\bibitem[{Varma {et~al.}(2019{\natexlab{a}})Varma, Field, Scheel, Blackman,
  Gerosa, Stein, Kidder, \& Pfeiffer}]{Varma-Vijay-PhysRevResearch.1.033015}
Varma, V., Field, S.~E., Scheel, M.~A., {et~al.} 2019{\natexlab{a}}, Phys. Rev.
  Research, 1, 033015

\bibitem[{Varma {et~al.}(2019{\natexlab{b}})Varma, Gerosa, Stein, H\'ebert, \&
  Zhang}]{Varma-Vijay-PhysRevLett.122.011101}
Varma, V., Gerosa, D., Stein, L.~C., H\'ebert, F. m.~c., \& Zhang, H.
  2019{\natexlab{b}}, Phys. Rev. Lett., 122, 011101

\bibitem[{Varma {et~al.}(2020)Varma, Isi, \&
  Biscoveanu}]{Varma-Vijay-PhysRevLett.124.101104}
Varma, V., Isi, M., \& Biscoveanu, S. 2020, Phys. Rev. Lett., 124, 101104

\bibitem[{Yang {et~al.}(2019)Yang, Bartos, Gayathri, Ford, Haiman, Klimenko,
  Kocsis, M\'arka, M\'arka, McKernan, \&
  O'Shaughnessy}]{Yang-Y-PhysRevLett.123.181101}
Yang, Y., Bartos, I., Gayathri, V., {et~al.} 2019, Phys. Rev. Lett., 123,
  181101

\end{thebibliography}
%

\end{document}